\documentclass[12pt]{article}
\usepackage{amsmath}
\usepackage{graphicx}
\usepackage{enumerate}
\usepackage{enumitem}
\usepackage{hyperref} 

\usepackage{url} 

\usepackage{geometry}    
\usepackage{float}
\usepackage{wrapfig}
\usepackage{sidecap}								
\usepackage{amssymb}
\usepackage{amssymb} 
\usepackage{amsmath} 
\usepackage{lipsum}
\usepackage{array} 
\usepackage{caption}
\usepackage{subcaption}
\usepackage{booktabs,caption}
\usepackage[flushleft]{threeparttable}
\usepackage{setspace}

\usepackage [english]{babel}
\usepackage [autostyle, english = american]{csquotes}
\MakeOuterQuote{"}

\newcommand{\blind}{1}

\graphicspath{ {./images/} }
\begin{document}

\def\spacingset#1{\renewcommand{\baselinestretch}%
{#1}\small\normalsize} \spacingset{0}


\if1\blind
{
  \title{\bf The Epistemology behind Covariate Adjustment}
  \author{Grayson L. Baird \footnote{Corresponding author: 535 Eddy St. Providence, RI 02903 \texttt{Grayson\char`_Baird@Brown.edu}}  \\
    Brown University \\
    and \\\
    Stephen L. Bieber \\\
    University of Wyoming}
  \maketitle
} \fi

\if0\blind
{
  \bigskip
  \bigskip
  \bigskip
  \begin{center}
    {\LARGE\bf The Epistemology behind Covariate Adjustment}
\end{center}
  \medskip
} \fi

\bigskip
\begin{abstract}
It is often asserted that to control for the effects of confounders, one should include the confounding variables of concern in a statistical model as a covariate. Conversely, it is also asserted that control can only be concluded by design, where the results from an analysis can only be interpreted as evidence of an effect because the design controlled for the cause. To suggest otherwise is said to be a fallacy of {\textit{cum hoc ergo propter hoc}}. Obviously, these two assertions create a conundrum: How can the effect of confounder be controlled for with analysis instead of by design without committing {\textit{cum hoc ergo propter hoc}}? The present manuscript answers this conundrum. 

\end{abstract} 

\noindent%
\vfill

\newpage
\spacingset{1.9} 
\section{Introduction}
\label{sec:intro}

It is often asserted that to control for the effects of confounds, one should include the confounding variables of concern in a statistical model as a covariate\footnote{Here and throughout this manuscript, a covariate is defined exclusively as an additive term in a model that is not the regressor of interest.} (e.g., Wysocki, Lawson, Rhemtulla, 2022; von Elm et al., 2007). Conversely, it is also asserted that control can only be achieved by design, where the results from an analysis can only be interpreted as evidence of an effect because the design controlled for the cause; to suggest otherwise is said to be a fallacy of {\textit{cum hoc ergo propter hoc}}: "with this, therefore because of this" (Fischer, 1970, p. 167). Obviously, these two assertions create a conundrum: Can the effect of a confound be controlled for by analysis instead of by design without committing {\textit{cum hoc ergo propter hoc}}? The purpose of the present manuscript is to answer this conundrum. 

\section{Ex Post Facto Data}

When modeling data from an {\textit{Ex Post Facto}} design, a common concern is how to control for confounding variables that likely exist but, given the nature of the non-experimental (observational) data, were not controlled for by design. A common solution to this problem is to include the confounding variables of concern in a statistical model as covariates, a practice often referred to as covariate adjustment. Covariate adjustment is simply the removal of the shared correlation among a regressor of interest, the outcome, and one or more confounding regressors of concern (the mathematics behind this are provided in the next section). It is believed that the regressor of interest, when adjusted, therefore, represents the relationship between the regressor of interest and the outcome controlling for the confounding regressor or regressors of concern. 

For example, let us assume that {\textit{Y}} is an outcome of interest while X\textsubscript{1} is a regressor of interest and X\textsubscript{2} is another regressor, a confound of concern, resulting in the following model multiple regression model$:$ \textup{\^Y}=b\textsubscript{0}+b\textsubscript{1}X\textsubscript{1}+b\textsubscript{2}X\textsubscript{2}. Mathematically, when X\textsubscript{2} is placed in a model, the relationship between {\textit{Y}} and X\textsubscript{1} is adjusted (i.e., b\textsubscript{1.2}) when the shared correlation between X\textsubscript{1}, {\textit{Y}}, and X\textsubscript{2} is removed. But is control over a confound actually achieved with adjustment by removing this shared correlation?

To be clear, if a confound in question were a regressor in the model, then the model would be adjusted---the shared correlation from that confounding regressor would be removed, thus controlling for that confound. However, the belief that adjustment is evidence itself of control of a confound is a logical fallacy known as {\textit{affirming the consequent\footnote{or in psychology, the contingency symmetry bias (Imai et al., 2021).}}: here, if control of a confound is achieved, then the coefficient will be adjusted; therefore, an adjusted coefficient is evidence of a controlled confound. The logical fallacy is committed because adjustment is a necessary condition of control but not a sufficient condition of control (Darner, 1987). The reason for this is straightforward: as the shared correlation is removed from both regressors, this removal is indiscriminate to cause itself---did X\textsubscript{1} cause the shared correlation, did X\textsubscript{2} cause the shared correlation, or was the shared correlation caused by X\textsubscript{3}, which was not in the model\footnote{referred to as the causal-arrow ambiguity (Meehl, 1970).}? Because covariate adjustment removes all shared correlation {\textit{regardless of the cause}, as a consequence, variance caused by the confound, if any, as well as variance not caused by a confound ("artifact") will be removed jointly.

The belief that adjustment is evidence of control of a confound is an example of postdiction\footnote{or in psychology, hindsight bias (Fischhoff, 1975).}---an explanation after the fact, which relies on another logical fallacy known as {\textit{post hoc ergo propter hoc}}: after this, therefore, because of this (Fischer, 1970, p. 166). Here, after an adjustment occurs, it is, therefore, because of the confound. Conversely, if no adjustment occurs, it is, therefore, because there was no confound, which also relies on the logical fallacy of negative proof or confusing the absence of evidence with evidence of absence (Fischer, 1970, p. 47). The logical fallacy is committed because other counterfactuals cannot be considered, such as adjustment not occurring but the confound still existing (i.e., undercorrelation, due to poor measurement, Kahneman, 1965), or more aptly, adjustment occurring but not because of the intended confound. As noted in the preceding paragraph, the latter outcome is the direct threat to the verisimilitude of covariate adjustment as a means of confound control---although covariate adjustment {\textit{can}} control for confounds, the problem is it is impossible to determine the degree to which the removal is due to the confound, if any, and the degree to which the removal is not due to the confound but instead artifact. But why is this a problem?

If the shared correlation that is removed is solely the confound, then the adjusted model reflects the population without a confound, as described by Bring (1994). However, if the removal includes shared correlation that is both confound and artifact, then what population does this adjusted model now represent? What does the model represent if the removal includes no confound and only artifact? In both cases, the adjusted model reflects neither the original population nor even the intended adjusted population---a discrepancy between the intended adjusted population and the population actually represented by the adjustment---known as model bias. But can this model bias be prevented?

Wysocki, Lawson, and Rhemtulla (2022) explicitly acknowledge that incorrect model adjustment must introduce bias in estimation. They hold that to prevent such bias, one needs to justify covariate adjustment with theory. Indeed, as far back as 1957, Kempthorne warns
\begin{quote}
 “the adjustment of data should be based on knowledge of how the factor which is being adjusted for actually produces its effect. An arbitrarily chosen adjustment formula may produce bias rather than remove the systematic difference. It is this fact which tends to vitiate the uses of the analysis of covariance recommended in most books on the analysis of experiments.” (p. 284).
\end{quote}
Obviously, theory-based adjustment is better than blind adjustment, such as the default adjustment of demographic variables, but theory alone cannot differentiate between confound and artifact from the shared correlation, and believing so relies upon yet another logical fallacy---special pleading---the application of a double standard or "if X then Y, but not when it hurts my agrument". (Fischer, 1970, p. 110). Here, the logical fallacy is committed when it is believed that covariate adjustment can serve as evidence for control simply because doing so is justified by some theory.

In conclusion, if covariate adjustment only demonstrates that shared correlation was removed, not that the confound and only the confound was removed in that correlation, then the belief that covariate adjustment serves as evidence of control over a confound is explicitly {\textit{cum hoc ergo propter hoc}}. A similar account of this fallacy was well documented by Meehl (1970), who observed
\begin{quote}
"it makes no sense to speak of a correlation as “spurious” or “in need of correction” unless a possible
error in causal-theoretical interpretation is envisaged ... In every instance that I have come across in which the investigator felt it necessary to employ partial correlation, analysis of covariance ... to 'avoid the influence' of an alleged nuisance variable, the rationale of such a procedure lay in his wish to conclude with a causal-theoretical inference or, at least, a counterfactual conditional of some kind."(p.8)
\end{quote}
In summary, the logical fallacy is committed not because covariate adjustment is unable to control for confounds but rather because covariate adjustment as a means of control is not {\textit{falsifiable}}, and, as such, therefore, adjustment cannot serve as evidence of control of a confound. 

\section{Experimental Data}

In the previous section, epistemology alone demonstrated why covariate adjustment could not serve as evidence of control of confounds for data from an {\textit{Ex Post Facto}} design, therefore revealing that the conundrum posed is both easily resolved and an issue not of mathematics but rather the misinterpretation of mathematics. If covariate adjustment cannot serve as evidence of control for confounds, then what can? Design can control for confounds, both known and unknown, by way of randomization\footnote{restriction is another method of control (of a known confound) by way of design.}. Here, control of confounds is achieved because the process of randomization guarantees that any observed differences between randomized groups must be due to chance variation, save for that which is being experimentally manipulated. If randomization controls for confounds by design {\textit{a priori}} of observation, then what would covariate adjustment achieve {\textit{post hoc}} of observation? 

Nuisance factors are undesired sources of systematic variation (bias) and random variation (error) occurring in an experiment that ultimately affect the outcome of interest (Kirk, 2013). It is believed that when these nuisance factors have been measured, referred to as "nuisance variables," these nuisance variables, therefore, can be controlled for using covariate adjustment. For example, if an imbalance of confounding variables between randomized groups is observed, then it is believed that randomization either created bias (Egbewale, Lewis, and Sim, 2014) or at least failed to control for bias from these confounding variables (Kahan et al., 2014). Another popular justification for using covariate adjustment with randomized data is the belief that adjustment reduces error between the randomized groups, thereby increasing statistical efficiency (Kahan et al., 2014; Pocock et al., 2002).

\subsection{Post-randomization control of confounds}

The former belief, that covariate adjustment controls for bias that randomization either created or failed to control, is predicated on three assumptions. The first assumption is that if confound imbalance is observed (at any defined alpha level), then the observed imbalance is evidence of a true difference between randomized groups---a bias (e.g., more males in one group versus another). The second assumption is that this observed bias is, therefore, evidence that randomization failed to control for the confounding variables of concern---the point of randomization is to control for bias; thus, if bias resulted after randomization, then randomization failed. The third assumption is that because randomization failed to control for bias, then the resulting bias can be controlled for using {\textit{post hoc}} covariate adjustment. 

The first assumption is easily refuted using fundamental statistical inference theory---hypothesis testing. Any difference observed in confounding variables between randomized groups must be due to chance variation; therefore, inferring a true difference in confounds between the randomized groups---a bias---is, by definition, a Type I error (a false discovery error) because the null is known to be true by design, so rejecting the null must be in error (Begg, 1987). The second assumption is also easily refuted. Imbalance cannot serve as evidence of bias because imbalance is a property of randomness---if randomization guaranteed balance across all factors, then randomization would be non-random as there would be a systematic selection bias against imbalance. Although possible, it is very unlikely that randomization will result in balance across all confounds, measured or unmeasured. 

In fact, the belief that confound imbalance is evidence that randomization "failed" to control for bias is the logical fallacy known as the fallacy of regression---the misperception of random events as non-random (Tversky and Kahneman, 1974), illustrated by Edwards Deming's red bead experiment (see Boardman, 1994). Here, because confound balance was not achieved, therefore, randomization "failed" to control for confound bias. The fallacy is committed because the distribution of confounds between randomized groups is, by design, random, therefore, any resulting imbalance, which must also be random, cannot serve as evidence that randomization "failed". Relatedly, the smaller the sample size, the more likely that confound imbalance will be observed. Therefore, the belief that more imbalance is evidence of more failure, not more random variability due to smaller sample sizes, is the insensitivity-to-sample-size cognitive bias (Tversky and Kahneman, 1974)---the misperception that variability, in this case, the chance of imbalance, does not change with sample size; simply, randomization cannot "fail" at any sample size.

The third assumption is easily refuted when considering the case in which randomization resulted in a confound imbalance between randomized groups to the extreme as if randomization had not even taken place. The belief that {\textit{post hoc}} covariate adjustment could control for this confound imbalance falls victim to the same logical fallacies as if randomization had not taken place (i.e., the {\textit{Ex Post Facto} design). And if this is the case for the most extreme instance of confound imbalance, then it follows this would be the case for lesser degrees of imbalance, too.  

The solution to confound imbalance is obvious: replication---the random imbalance of variables, confound or not, observed for a given trial will balance out with several repeated trials---the random imbalance of variables will average out over the long run\footnote{another solution is by design, using balance block, urn, etc., though these solutions deviate from random and only control measured and specified variables.}. In this way, covariate adjustment in response to these random imbalances will vary from trial to trial, resulting in a game of adjustment Whac-A-Mole\textsuperscript{TM}; any unintended model bias that results from these adjustments can, therefore, be a source of replication discordance in the literature that would otherwise not occur had covariate adjustment not been used. Although this issue was touched upon theoretically and empirically by de Boer et al. (2015), their solution---using theory to guide covariate selection instead of random confound imbalance---rests upon logical fallacies outlined in the preceding section. 

Finally, some acknowledge that confound imbalances are random; however, they still hold that not using covariate adjustment will still result in bias (Sherry et al., 2023). Here, it is believed that if confounding variables are selected with theory and {\textit{a priori}} to randomization, then adjustment will result in decreased bias (note: why randomization failed to control for this bias is never explored). This line of reasoning is an improvement to blind covariate adjustment as a reaction to random confound imbalance, though again, it still falls victim to the same logical fallacies presented in the previous section and, in particular, a special pleading.

\subsection{Post-randomization control of error}

The latter belief, that covariate adjustment controls for (or reduces) error, is based on the knowledge that because covariate adjustment can explain some variation that occurs in outcomes between randomized groups, covariate adjustment reduces unexplained variance which in turn, increases statistical efficiency. As Kahan et al. (2014) note, the greater the error that is "reduced depends on the correlation between the covariates and the outcome; the higher the correlation, the larger the increase in power". Although this certainly can be true, this reduction in unexplained variance can also come at the cost of increased bias. The present section demonstrates how covariate adjustment may reduce unaccounted-for variance but at the cost of introducing bias into the model, where the inference and estimation of the adjusted model become divorced from the original population it was intended to represent.

To demonstrate this, we will compare data that represent two possibilities: data in which a nuisance variable does not correlate with the regressor of interest but does correlate with the outcome and data in which a nuisance variable correlates with the regressor of interest and the outcome jointly. For simplicity and without loss of generality, the two-regressor model is used, assuming randomization took place but without any manipulation. This is done as the issue at hand deals with the "correlation between the covariates and the outcome" (Kahan et al., 2014, p.2); thus, we will examine how adjustment reduces the unaccounted-for variance (error) between the covariates and outcome by explaining "nuisance" variation in isolation of any manipulation. Thus, we show how bias is introduced with covariate adjustment, even without experimental manipulation. 

We will first define covariate adjustment by demonstrating the process by which covariate adjustment is achieved for the multiple linear regression model (MLR).  Each step of this demonstration will provide graphical evidence, using Ballantine Venn diagram (Cohen and Cohen, 1975; Baird and Bieber, 2023) representing a 3-variable situation, mathematical equations, and empirical evidence, using data from a simulation (code found in Appendix A).

\spacingset{1.9} 
\subsubsection{No Covariate Adjustment}
\label{sec:intro}

Again, let us assume that {\textit{Y}} is an outcome of interest while X\textsubscript{1} is a regressor of interest and X\textsubscript{2} is another regressor, the covariate, both of which relate with {\textit{Y}}. This is illustrated conceptually in Figure 1 as a Venn diagram, where X\textsubscript{1} and X\textsubscript{2}, represented as areas 1 and 2, respectively, are independent of each other and both relate with Y, and the remainder is represented as area 3, denoted as Y\textsubscript{123}.  This is also represented as a covariance matrix in Figure 2, defined as {\bf COV1 = [Y X\textsubscript{1} X\textsubscript{2}}]. When both regressors, X\textsubscript{1} and X\textsubscript{2}, are uncorrelated such that COV1(X\textsubscript{1},X\textsubscript{2})=0.00 and are placed together into the model, defined as \textup{\^Y}=b\textsubscript{0}+b\textsubscript{1}X\textsubscript{1}+b\textsubscript{2}X\textsubscript{2}, the regressors need not be adjusted, as seen in Equations (1) and (2).

\begin{figure}
\centering
 \small
\begin{minipage}[b]{.30\textwidth}
\includegraphics[width=\textwidth]{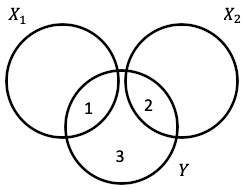}
\caption{Y\textsubscript{123}} \label{first}
\end{minipage}\hfill
\begin{minipage}[b]{.30\textwidth}
\centering
\large
\label{M}
\(
COV1=
\begin{pmatrix}
2.00&0.65&0.65\\
0.65&0.60&0.00\\
0.65&0.00&0.60\\
\end{pmatrix}
\)
\caption{Matrix}\label{second}
\end{minipage}
\end{figure}

\begin{equation} 
\begin{split}
{\textup{\^{X}}_{1}}=({\textup{c}_{0}}+{\textup{c}_{2}}{\textup{X}_{2}}) = 0\\
{\textup{X}_{1.2}}={\textup{X}_{1}}-{\textup{\^{X}}_{1}} = {\textup{X}_{1}}
\end{split}
\end{equation} yielding a residualized regressor X\textsubscript{1.2} and 

\begin{equation} 
\begin{split}
{\textup{\^{X}}_{2}}=({\textup{c}_{0}}+{\textup{c}_{1}}{\textup{X}_{1}}) = 0\\
{\textup{X}_{2.1}}={\textup{X}_{2}}-{\textup{\^{X}}_{2}} = {\textup{X}_{2}}
\end{split}
\end{equation} yielding a residualized regressor X\textsubscript{2.1}. In other words, if there is no shared correlation between regressors and {\textit{Y}}, then there is no need for covariate adjustment and thus, both regressors simply reduced to themselves, X\textsubscript{1} and X\textsubscript{2}, respectively.

These regressors are then used to calculate the MLR coefficients, seen in Equations (3) and (4).

\begin{equation} 
\begin{split}
{\textup{b}_{1.2}}=\frac {\sum{\textup{X}_{1.2}}\cdot Y} {\sum({\textup{X}_{1.2}})\textsuperscript{2}}=
{\textup{b}_{1}}=\frac {\sum{\textup{X}_{1}}\cdot Y} {\sum({\textup{X}_{1}})\textsuperscript{2}}\\
 =\frac{6606.337} {6014.009}= 1.099
\end{split}
\end{equation}

\begin{equation} 
\begin{split}
{\textup{b}_{2.1}}=\frac{\sum{\textup{X}_{2.1}}\cdot Y} {\sum({\textup{X}_{2.1}})\textsuperscript{2}} =
{\textup{b}_{2}}=\frac{\sum{\textup{X}_{2}}\cdot Y} {\sum({\textup{X}_{2}})\textsuperscript{2}} \\
=\frac{6554.81} {6037.42}=1.090
\end{split}
\end{equation}

These coefficients are then used to calculate the sums of squares of the model, as seen in Equations (5) and (6):

\begin{equation} 
\begin{split}
SS(X\textsubscript{1}\vert X\textsubscript{2})={\textup{b}_{1.2}}\sum{\textup{X}_{1.2}}\cdot Y =
SS(X\textsubscript{1})={\textup{b}_{1}}\sum{\textup{X}_{1}}\cdot Y \\
=1.099 \cdot 6606.34=7257.01
\end{split}
\end{equation}

\begin{equation} 
\begin{split}
SS(X\textsubscript{2}\vert X\textsubscript{1})={\textup{b}_{2.1}}\sum{\textup{X}_{2.1}}\cdot Y =
SS(X\textsubscript{2})={\textup{b}_{2}}\sum{\textup{X}_{2}}\cdot Y\\
=1.090 \cdot 6554.81=7116.53 
\end{split}
\end{equation}

and the sums of squares of the error of the model, as seen in Equation (7):

\begin{equation} 
SS(E)=\sum({\textup{Y}}-{\textup{\^Y}})\textsuperscript{2}=5760.98 \end{equation}

When the sums of squares for the regressors and the error are then added together, the total sums of squares is the following

\begin{equation} 
\begin{split}
SS({\textup{Total}})=({\textup{b}_{1.2}}\sum{\textup{X}_{1.2}}\cdot Y)+({\textup{b}_{2.1}}\sum{\textup{X}_{2.1}}\cdot Y) + \sum({\textup{Y}}-{\textup{\^Y}})\textsuperscript{2} \\
=({\textup{b}_{1}}\sum{\textup{X}_{1}}\cdot Y)+({\textup{b}_{2}}\sum{\textup{X}_{2}}\cdot Y) + \sum({\textup{Y}}-{\textup{\^Y}})\textsuperscript{2} \\= 7257.01+7116.53+5760.98=20134.51,
\end{split}
\end{equation}

which is the same value\footnote{empirically, the small difference is due to simulation and round-off error} as the sums of squares of {\textit{Y}}, as seen in Equation (9)

\begin{equation}
\sum{\textup{Y}}\textsuperscript{2} - \frac{(\sum{\textup{Y}})\textsuperscript{2}} {N}=20150.29-\frac{15506.23} {10000} =20148.74.\end{equation}

For reference, again, this is illustrated conceptually in Figure 1, where the outcome Y is comprised of the unique correlation with X\textsubscript{1} (area 1), X\textsubscript{2} (area 2), and error, (area 3), and will be denoted as {Y\textsubscript{123}}. The regression model, conceptualized in Figure 3, is denoted as \^{Y\textsubscript{12.3}}.

\begin{figure}
\centering
 \small
\begin{minipage}[b]{.30\textwidth}
\includegraphics[width=\textwidth]{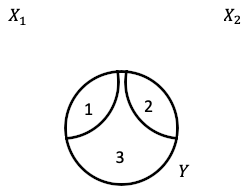}
\caption{The regression model, \^{Y\textsubscript{12.3}}, or the outcome, Y\textsubscript{123} }\label{first}
\end{minipage}\hfill
\end{figure}

Figure 3 also conceptually represents the variance accounted for by the model, where area 1 represents SS(X\textsubscript{1}), area 2 represents SS(X\textsubscript{2}), and area 3 represents SS(E), which added together represents SS(Total). This value is quantified by the calculation of the coefficient of determination or R\textsuperscript{2}: 

\begin{equation} 
{\textup{R}}\textsuperscript{2} = \frac {SS(Total)-SS(E)} {SS(Total)} = \frac {20134.51-5760.98} {20134.51} = .71 \end{equation} or equivalently, 

\begin{equation}
\begin{split}
{\textup{R}}\textsuperscript{2} = 
\frac {SS(X\textsubscript{1}\vert X\textsubscript{2}) + SS(X\textsubscript{2}\vert X\textsubscript{1}) } {SS(X\textsubscript{1}\vert X\textsubscript{1}) + SS(X\textsubscript{2}\vert X\textsubscript{1}) + SS(E) } =
\frac {SS(X\textsubscript{1}) + SS(X\textsubscript{2}) } {SS(X\textsubscript{1}) + SS(X\textsubscript{2}) + SS(E) } \\
 = \frac {7257.01+7116.53} {7257.01+7116.53+5760.98}=.71. 
 \end{split}
\end{equation} This can be confirmed by comparing the variance components of the model, ${\sigma}_{\hat{Y}_{12.3}}^2$, with the variance components of the model ${\sigma}_{\hat{Y}_{12.3}}^2$ and error ${\sigma}_{e}\textsuperscript{2}$, or ${\sigma}_{Y_{123}}^2$, as seen in Equation (12):

\begin{equation} 
{\textup{R}}\textsuperscript{2} = \frac { {\sigma}_{\hat{Y}_{12.3}}^2 } {  {\sigma}_{Y_{123}}^2 } = \frac {1.44} {2.02} = .71  \end{equation}

where

\begin{equation}
\begin{split}
{\sigma}_{\hat{Y}_{12.3}}^2  = 
{b}_{1.2}\textsuperscript{2}  {\sigma}_{X1.2}\textsuperscript{2} + {b}_{2.1}\textsuperscript{2}  {\sigma}_{X2.1}\textsuperscript{2} =
{b}_{1}\textsuperscript{2}  {\sigma}_{X1}\textsuperscript{2} + {b}_{2}\textsuperscript{2}  {\sigma}_{X2}\textsuperscript{2} 
\\
=(1.099 \cdot 0.60)+(1.090 \cdot 0.60) =1.44 
\end{split}
\end{equation} is the variance components of the model (areas 1 and 2 in Figure 3) and,
\begin{equation}
\begin{split}
{\sigma}_{Y_{123}}^2 = {b}_{1.2}\textsuperscript{2}  {\sigma}_{X1.2}\textsuperscript{2} + {b}_{2.1}\textsuperscript{2}  {\sigma}_{X2.1}\textsuperscript{2} + {\sigma}_{e}\textsuperscript{2}=
{b}_{1}\textsuperscript{2}  {\sigma}_{X1}\textsuperscript{2} + {b}_{2}\textsuperscript{2}  {\sigma}_{X2}\textsuperscript{2} + {\sigma}_{e}\textsuperscript{2} \\
= (1.099 \cdot 0.60)+(1.090 \cdot 0.60)+0.576  =2.02,
\end{split}
\end{equation} the total variance components of the model and error (areas 1, 2, and 3 in Figure 3) which is also the same value as the variance of {\textit{Y}}:

\begin{equation}
{\sigma}_{Y}\textsuperscript{2} ={\sigma}_{Y_{123}}^2 = 2.02. \end{equation}

The variance accounted for by the model can also be confirmed by summing the squared standardized coefficients (area 1 and 2 in Figure 3), as seen in Equation (16):

\begin{equation}
\begin{split}
{\textup{R}}\textsuperscript{2} = {\beta}_{1.2}\textsuperscript{2} + {\beta}_{2.1}\textsuperscript{2}=
{\beta}_{1}\textsuperscript{2} + {\beta}_{2}\textsuperscript{2} \\
= 0.36+ 0.35 = .71 
\end{split}
\end{equation}

where

 \begin{equation}
 \begin{split}
{\beta}_{1.2}=\frac { {b}_{1.2} \cdot {\sigma}_{X1.2}} {{\sqrt{{\sigma}_{Y_{123}}^2}}}  =
{\beta}_{1}=\frac { {b}_{1} \cdot {\sigma}_{X1}} {{\sqrt{{\sigma}_{Y}^2}}}= \\
\frac {1.099 \cdot 0.78} {\sqrt{{1.42}}}  =0.60
\end{split}
\end{equation}

and 

\begin{equation}
\begin{split}
{\beta}_{2.1}=\frac { {b}_{2.1} \cdot {\sigma}_{X2.1}} {{\sqrt{{\sigma}_{Y_{123}}^2}}} =
{\beta}_{2}=\frac { {b}_{2} \cdot {\sigma}_{X2}} {{\sqrt{{\sigma}_{Y}^2}}} =\\
 \frac {1.090 \cdot 0.78} {\sqrt{{1.42}}}  =0.60.
 \end{split}
 \end{equation}

As illustrated in Figure 3 (area 2) and made obvious in Equations 12-14, the inclusion of the covariate reduced the value of the unaccounted-for variance, thereby increasing the statistical efficiency of the model. It is important to note that while the inclusion of the covariate reduced the value of the error, the sum of the variance components of the model still equal the variance of the outcome as made evident with Equation 15 and illustrated when comparing Figures 1 with 3. 
 
\spacingset{1.9} 
\subsubsection{Covariate Adjustment}
\label{sec:intro}

Now let us consider when the regressors are not independent. This is illustrated conceptually in Figure 4, where outcome Y is comprised of the unique correlation of X\textsubscript{1} (area 1), X\textsubscript{2} (area 2), the shared correlation of X\textsubscript{1} and X\textsubscript{2} (area 4), and error (area 3), which will be denoted as Y\textsubscript{1234}. This is also represented as a covariance matrix in Figure 5, defined as {\bf COV2 = [Y X\textsubscript{1} X\textsubscript{2}}]. 

\begin{figure}
\centering
 \small
\begin{minipage}[b]{.25\textwidth}
\includegraphics[width=\textwidth]{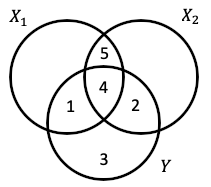}
\caption{Y\textsubscript{1234} }\label{first}
\end{minipage}\hfill
\begin{minipage}[b]{.30\textwidth}
\centering
\large
\label{M}
\(
COV2=
\begin{pmatrix}
2.00&0.65&0.65\\
0.65&0.60&0.25\\
0.65&0.25&0.60\\
\end{pmatrix}
\)
\caption{Matrix Structure}\label{second}
\end{minipage}
\end{figure}

A close inspection reveals Equations (19-22) are identical to Equations (1-4), but the results differ dramatically when X\textsubscript{1} and X\textsubscript{2} share correlation with {\textit{Y}}:

\begin{equation} 
\begin{split}
{\textup{\^{X}}_{1}}=({\textup{c}_{0}}+{\textup{c}_{2}}{\textup{X}_{2}}) \\
{\textup{X}_{1.2}}={\textup{X}_{1}}-{\textup{\^{X}}_{1}}
\end{split}
\end{equation} what remains in X\textsubscript{1} and Y (area 1 and 4) after removing X\textsubscript{2} (removing area 2 and 4) is a residualized regressor, X\textsubscript{1.2} (area 1) and

\begin{equation} 
\begin{split}
{\textup{\^{X}}_{2}}=({\textup{c}_{0}}+{\textup{c}_{1}}{\textup{X}_{1}}) \\
{\textup{X}_{2.1}}={\textup{X}_{2}}-{\textup{\^{X}}_{2}}
\end{split}\end{equation} what remains in X\textsubscript{2} and Y (area 2 and 4) after removing X\textsubscript{1} (removing area 1 and 4) is a residualized regressor, X\textsubscript{2.1} (area 2). These adjusted (residualized) regressors are then used to calculate the adjusted MLR coefficients in the following manner\footnote{Frisch-Waugh-Lovell theorem; see Yule (1907), Frisch and Waugh (1933), Lovell (1963).}:

\begin{equation} 
{\textup{b}_{1.2}}=\frac {\sum{\textup{X}_{1.2}}\cdot Y} {\sum({\textup{X}_{1.2}})\textsuperscript{2}}=\frac{3871.54} {4984.37}=0.78\end{equation}

\begin{equation}  
{\textup{b}_{2.1}}=\frac{\sum{\textup{X}_{2.1}}\cdot Y} {\sum({\textup{X}_{2.1}})\textsuperscript{2}} =\frac{3881.04} {4964.31}= 0.78\end{equation} 

Because X\textsubscript{1} and X\textsubscript{2} are correlated with each other and {\textit{Y}}, the shared correlation has been removed via Equations (19-22), producing adjusted coefficients b\textsubscript{1.2} and b\textsubscript{2.1}, represented as areas 1 and 2 in Figure 6, (cf. instead of non-adjusted coefficients b\textsubscript{1} and b\textsubscript{2}, represented in area area 1 and 2 in Figure 3). The model’s sums of squares\footnote{Equations 19-22 represent the Type III or unique sums of squares. To the authors' knowledge, this is the first place where the derivation for Type III sums of squares has been provided for multiple linear regression.} are then the following:

\begin{equation} 
SS(X\textsubscript{1}\vert X\textsubscript{2})={\textup{b}_{1.2}}\sum{\textup{X}_{1.2}}\cdot Y=0.78 \cdot 3871.54=3007.17\end{equation}

\begin{equation} 
SS(X\textsubscript{2}\vert X\textsubscript{1})={\textup{b}_{2.1}}\sum{\textup{X}_{2.1}}\cdot Y=0.78 \cdot 3881.04=3034.14.\end{equation}

When the sums of squares for the regressors and the error are then added together, the total sums of squares is the following

\begin{equation}
\begin{split} 
SS({\textup{Total}})=({\textup{b}_{1.2}}\sum{\textup{X}_{1.2}}\cdot Y)+({\textup{b}_{2.1}}\sum{\textup{X}_{2.1}}\cdot Y) +  \sum({\text{Y}}-{\textup{\^Y}})\textsuperscript{2} \\=3007.17+3034.14+9843.36=15884.67 
\end{split}
\end{equation}
which is no longer the same value as the sums of squares of {\textit{Y}},

\begin{equation}
\sum{\textup{Y}}\textsuperscript{2} - \frac{(\sum{\textup{Y}})\textsuperscript{2}} {N}=20150.29-\frac{15506.23} {10000} = 20148.74
\end{equation}
a difference of 4264.07, or a 21 percent reduction in overall variance of {\textit{Y}}. If the shared variance is not contained in the sums of squares for either regressor, then the shared variance cannot be contained in either the model sums of squares or the total sums of squares (see Woolf, 1951; Baird and Bieber, 2023). This is because the shared correlation was removed not once but twice (Kerlinger and Pedhazur 1973, p. 46), making the shared correlation absent from the model (Newton and  Spurrell, 1967, p. 57; Kennedy, 1982, p. 63) and the variance for which it accounts. As before, Figure 6 conceptually represents the variance accounted for by the model, where area 1 represents $SS(X\textsubscript{1}\vert X\textsubscript{2})$, area 2 represents $SS(X\textsubscript{2}\vert X\textsubscript{1})$, and area 3 represents SS(E).

\begin{figure}
\centering
 \small
\begin{minipage}[b]{.25\textwidth}
\includegraphics[width=\textwidth]{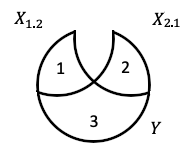}
\caption{Regression model, \^{Y\textsubscript{12.34}}, or Y\textsubscript{123.4}  }\label{first}
\end{minipage}\hfill
\end{figure}

Just as in Figure 3, Figure 6 reveals that the error is reduced when placing X\textsubscript{2} in the model as a covariate. However, the relationship between X\textsubscript{1} and {\textit{Y}} is now adjusted because the shared correlation between X\textsubscript{2} and X\textsubscript{1} has been removed (area 4 in Figure 4), which reveals the shared correlation from X\textsubscript{1} and X\textsubscript{2} (area 4) has been removed from the outcome, denoted as Y\textsubscript{123.4}. Likewise, the adjusted model is denoted as \^Y\textsubscript{12.34}. 

Identical with Equations (10,11,12,16), Equations (27,28,29,33) then become the basis for calculating the R\textsuperscript{2} of the adjusted model:

\begin{equation} 
{R}\textsuperscript{2} = \frac {\textup{Total}-SS(E)} {\textup{Total}}= \frac {15884.67-9843.363} {15884.67} =.38  \end{equation}

or equivalently 

\begin{equation}
{R}\textsuperscript{2}= \frac {SS(X\textsubscript{1}\vert X\textsubscript{2}) + SS(X\textsubscript{2}\vert X\textsubscript{1}) } {SS(X\textsubscript{1}\vert X\textsubscript{2}) + SS(X\textsubscript{2}\vert X\textsubscript{1}) + SS(E) }= \frac {3007.17+3034.14} {3007.17+3034.14+9843.36} =.38
 \end{equation}

This can be confirmed by comparing the variance components of the adjusted model, ${\sigma}_{\hat{Y}_{12.34}}^2$, with the variance components of the adjusted model ${\sigma}_{\hat{Y}_{12.34}}^2$ and error ${\sigma}_{e}\textsuperscript{2}$, or ${\sigma}_{Y_{123.4}}^2$, as seen in Equation (29):

\begin{equation} 
{\textup{R}}\textsuperscript{2} = \frac { {\sigma}_{\hat{Y}_{12.34}}^2 } {  {\sigma}_{Y_{123.4}}^2 } = \frac {0.60} {1.59} = .38 \end{equation} 

where 

\begin{equation}
{\sigma}_{\hat{Y}_{12.34}}^2 = {b}_{1.2}\textsuperscript{2}  {\sigma}_{X1.2}\textsuperscript{2} + {b}_{2.1}\textsuperscript{2}  {\sigma}_{X2.1}\textsuperscript{2}=(0.60 \cdot 0.50) + (0.61 \cdot 0.50) =0.60 \end{equation} is the variance components of the adjusted model (areas 1 and 2 in Figure 6) and,

\begin{equation}
{\sigma}_{Y_{123.4}}^2 = {b}_{1.2}\textsuperscript{2}  {\sigma}_{X1.2}\textsuperscript{2} + {b}_{2.1}\textsuperscript{2}  {\sigma}_{X2.1}\textsuperscript{2} + {\sigma}_{e}\textsuperscript{2}=(0.60 \cdot 0.50) + (0.61 \cdot 0.50) + 0.98  =1.59 \end{equation} is the total variance components of the adjusted model and error (areas 1, 2, and 3 in Figure 6) which is no longer the same value as the variance of {\textit{Y}}:

\begin{equation}
{\sigma}_{Y}\textsuperscript{2} =2.02 \neq {\sigma}_{Y_{123.4}}^2 = 1.59 \end{equation}.

As before, the variance accounted for by the model can also be confirmed by summing the squared adjusted standardized coefficients, as seen in Equation (33):

\begin{equation}
{R}\textsuperscript{2}  = {\beta}_{1.2}\textsuperscript{2} + {\beta}_{2.1}\textsuperscript{2}= 0.19+ 0.19 = 0.38 \end{equation}

where
 
\begin{equation}
{\beta}_{1.2}=\frac {{b}_{1.2} {\sigma}_{X1.2}} {{\sqrt{{\sigma}_{Y_{123.4}}^2}}} =\frac {0.78 \cdot 0.71} {\sqrt{{1.26}}}=0.44 \end{equation}

and 

\begin{equation}
{\beta}_{2.1}=\frac {{b}_{2.1} {\sigma}_{X2.1}} {{\sqrt{{\sigma}_{Y_{123.4}}^2}}}= \frac {0.78 \cdot 0.71} {\sqrt{{1.26}}} =0.44\end{equation}.

As illustrated in Figure 6 (area 2) and made obvious in Equations 29-31, the inclusion of the covariate reduced the value of the unaccounted-for variance, thereby increasing the statistical efficiency of the model. However, unlike the previous example, it is important to note that while the inclusion of the covariate reduced the value of the error, the sum of the variance components of the model no longer equals the variance of the outcome, as made evident with Equation 32 and illustrated comparing Figure 4 with 6.

\spacingset{1.9} 
\section{Discussion}
\label{sec:intro}

Although Equations 16-18 and 33-35 are identical, only the former refers to the model's accounting of variance for the original outcome, {\textit{Y}}, while the latter refers to the adjusted outcome {\textit{Y}}\textsubscript{123.4}. That is, when there is no covariate adjustment, the model coefficients, model variance components, and model sums of squares, along with the corresponding error, together represent the original population, {\textit{Y}. Conversely, when covariate adjustment takes place, the "additive" model no longer adds to the original variance of {\textit{Y}; instead, the model represents a population without the shared correlation, an "adjusted" population, {\textit{Y}\textsubscript{123.4}. But what does the discrepancy between the original population and this adjusted population represent? 

Clearly, the adjusted population {\textit{ipso facto}} no longer represents the original population. The reason for the adjustment is simple: when regressors are correlated with each other and the outcome variable, this shared correlation is removed entirely, including from the outcome itself. But this reduction in error affects the {\textit{interpretation}} of the model$:$ when adjusting for X\textsubscript{2}, for every increase in $\sigma$\textsubscript{X1.2}, a $\beta$\textsubscript{1.2} change in ${\sigma}_{Y_{123.4}}$ occurs, which is clear from the variance components in Equation (31) and illustrated in Figure 6. Now consider the interpretation when no covariate adjustment has taken place$:$ for every increase in $\sigma$\textsubscript{X1}, a $\beta$\textsubscript{1} change in $\sigma$\textsubscript{Y} occurs, which is clear from the variance components in Equation (14) and illustrated in Figure 3. This distinction in interpretation was perhaps first made explicit by Bring (1994, p. 211), but the fact that the model no longer represents the original population is hidden by the common {\textit{misinterpretation}} of the adjusted model$:$ when adjusting for X\textsubscript{2}, for every increase in $\sigma$\textsubscript{X1}, a $\beta$\textsubscript{1.2} change in ${\sigma}_{Y}$ occurs, a subtle but important bait-and-switch, which hides the discrepancy between the original population and what the adjustment model represents. 

As before, the issue is not one of mathematics but instead the misinterpretation of mathematics. Covariate adjustment, symbolized by the lacuna (area 4) when comparing Figure 6 with Figure 4, illustrates the successful reduction of unaccounted-for variance in the model, but it also illustrates the cost of this reduction: the model is no longer representative of the original population. Conversely, if a covariate is included in the model but no adjustment takes place, then the model remains representative of the original population, even though the covariate reduced the unaccounted-for variance, symbolized by comparing Figures 1 and 3. Therefore, the cost of increased statistical efficiency using a covariate is an increase in bias (removal of area 4) but only when adjustment results from the inclusion of the covariate. This is likely why, over 70 years ago, when Woolf first discovered the effects of adjustment on the variance components, he concluded, "there seems to be an enormous deficit in the balance sheet. The paradox, of course, is easily resolved. It is essential for the validity of an analysis of variance that the various components shall be uncorrelated or independent" (1951, p.113).  

At this point, the question is clear: why use covariate adjustment? When control of confounds or nuisance variables is not possible by design, it could be argued that covariate adjustment must then be used. This decision is predicated on {\textit{faith}}---faith that adjustment reduces the intended confound without simultaneously introducing more bias or faith that although adjustment may lead to increased statistical efficiency, this efficiency is not also creating a discrepancy with which the model no longer reflects the population it is supposed to represent---again bias. Another option is to use the unadjusted model with replication. This decision is predicated on parsimony, where there is no threat of introducing more bias than what already exists and with the knowledge that the inference and estimation of the model reflect the same population that was originally observed and can be directly defined by the design.

Faith will likely prevail because the incentives are aligned, a possible result of Goodhart's Law (Rodamar, 2018). With the inherent pressures to achieve statistical significance with as little data and replication as possible, the promise of the former approach---control of confounds or reduction of error---is perceived to outweigh the likely cost---the introduction of bias---because the consequences of bias cannot be quantified. The impulse to use modeling to achieve what can only be achieved by design is hardly a new phenomenon. Now, over 90 years ago, Nikola Tesla observed: "Today's scientists have substituted mathematics for experiments, and they wander off through equation after equation, and eventually build a structure which has no relation to reality" (1934, p. 117-118). Only four short years later, in his "The Misuse of Statistics," Cohen (1938) cautioned, "...unjustifiable conclusions may be drawn, and with instances where incorrect inferences are drawn from valid data. The trained statistician absorbed in the improvement of technical procedure grows careless in interpretation..." (p. 660). Likely, these traditions will continue. 

\clearpage

\section{Appendix A: Simulation Study, R Code}
\begin{singlespace}
\begin{verbatim}

#####R Code

#### Covariance Matrix (COV.1), Figure 2, Uncorrelated by design
nobs = 10000
set.seed(2001)
M = matrix(c(2.00, 0.65, 0.65,
             0.65, 0.60, 0.00,
             0.65, 0.00, 0.60), nrow=3, ncol=3)

#### Covariance Matrix (COV.2), Figure 5, Shared Correlation by design
nobs = 10000
set.seed(2001)
M = matrix(c(2.00, 0.65, 0.65,
             0.65, 0.60, 0.25,
             0.65, 0.25, 0.60), nrow=3, ncol=3)
             
### Cholesky decomposition
L = chol(M)
nvars = dim(L)[1]
t(L)
t(L) %*% L
r = t(L) %*% matrix(rnorm(nvars*nobs), nrow=nvars, ncol=nobs)
r = t(r)
SimM = as.data.frame(r)
names(SimM) = c("Y", "X1", "X2")
Y=SimM$Y
X1=SimM$X1
X2=SimM$X2

##Model fit
MLR = lm(Y~ X1 + X2, SimM)
resid_Y=resid(MLR)

#Equation 1/19, Residualized Predictor
residX1 = lm(X1~ X2, SimM)
X1.2=resid(residX1)

#Equation 2/20, Residualized Predictor
residX2 = lm(X2~ X1, SimM)
X2.1=resid(residX2)

#Equation 3/21, Unstandardized Coefficient
b1.2=sum((X1.2*Y))/sum(X1.2^2)
b1.2

#Equation 4/22, Unstandardized Coefficient
b2.1=sum((X2.1*Y))/sum(X2.1^2)
b2.1

#Equation 5/23, Type III Sums of Squares
SSb1.2=b1.2*sum(X1.2*Y)
SSb1.2

#Equation 6/24, Type III Sums of Squares
SSb2.1=b2.1*sum(X2.1*Y)
SSb2.1

#Equation 7, Error Sums of Squares
SSE = sum(resid(MLR)^2)
SSE

#Equation 8/25, Total Sums of Squares by parts
SSTotal.12=SSb1.2+SSb2.1+SSE
SSTotal.12

#Equation 9/26, Sums of Squares Y
SSTotal= sum(Y^2) - ((sum(Y)^(2))/10000)
SSTotal

#Equation 10/27 , Coefficient of determination
R2=(SSTotal.12-SSE)/SSTotal.12
R2

#Equation 11/28 , Coefficient of determination
R2=(SSb1.2+SSb2.1)/(SSb1.2+SSb2.1+SSE)
R2

#Equation 13/30 Variance components of model
Sig_Yp=(((b1.2)^2)*var(X1.2)) + (((b2.1)^2)*var(X2.1))
Sig_Yp

#Equation 14/31 Variance components of model and error
Sig_Y=(((b1.2)^2)*var(X1.2)) + (((b2.1)^2)*var(X2.1))+var(resid_Y)
Sig_Y

#Equation 12/29 , Coefficient of determination
R2=Sig_Yp/Sig_Y
R2

#Equation 15/32, Variance of Y
Var_Y=var(Y)
Var_Y

#Equation 17/34  Standardized Coefficient
B1.2= (b1.2*(sd(X1.2))/sqrt(Sig_Y))
B1.2

#Equation 18/35 Standardized Coefficient
B2.1= (b2.1*(sd(X2.1))/sqrt(Sig_Y))
B2.1

#Equation 16/33, Coefficient of determination
R2=(B1.2^2)+(B2.1^2)
R2

###############################
#Type III Sums of Squares Check
library("car")
Anova(MLR, type="III")

##Reference to shared correlation that was removed removed
Area4=var(Y)-Sig_Y
Area4



\end{verbatim}
\end{singlespace}

\newpage
 
\section{References}

\begin{enumerate}[leftmargin=!,labelindent=5pt,itemindent=-15pt]

\item Baird, G. L., and Bieber, S. L. (2016). The Goldilocks dilemma: Impacts of multicollinearity-a comparison of simple linear regression, multiple regression, and ordered variable regression models. {\textit{Journal of Modern Applied Statistical Methods, 15(1),}} 18.

\item Baird, G. L., and Bieber, S. L. (2020). Sampling the porridge: A comparison of ordered variable regression with F and R\textsuperscript{2} and multiple linear regression with corrected F and R\textsuperscript{2} in the presence of multicollinearity.  {\textit{Journal of Modern Applied Statistical Methods, 18(1)}}, 11.   

\item Baird, G. L., and Bieber, S. L. (2023). Revisiting the conceptualization of multiple linear regression. {\textit{arXiv preprint}} arXiv:2302.06464.

\item Begg, C. B. (1990). Significance tests of covariate imbalance in clinical trials. {\textit{Controlled Clinical Trials}}, 11(4), 223-225.
   
\item Bring, J. (1994). How to standardize regression coefficients, {\textit{The American Statistician, 48(3),  209-213}}.

\item Boardman, T. J. (1994). The statistician who changed the world: W. Edwards Deming, 1900–1993. {\textit{The American Statistician}}, 48(3), 179-187.

\item Cohen, J. and P. Cohen. {\textit{Applied multiple regression/correlation analysis for the behavioral
sciences}}. Hillside. N.J.: Lawrence Erlbaum Associates. 1975. 

\item Cohen, J. B. (1938). The misuse of statistics. {\textit{Journal of the American Statistical Association}}, 33(204), 657-674.

\item Darner, T. E. (1987). {\textit{Attacking faulty reasoning}}. Belmonth, CA: Wadsworth.

\item De Boer, M. R., Waterlander, W. E., Kuijper, L. D., Steenhuis, I. H., and Twisk, J. W. (2015). Testing for baseline differences in randomized controlled trials: An unhealthy research behavior that is hard to eradicate. {\textit{International Journal of Behavioral Nutrition and Physical Activity}}, 12, 1-8.

\item Egbewale, B. E., Lewis, M., and Sim, J. (2014). Bias, precision and statistical power of analysis of covariance in the analysis of randomized trials with baseline imbalance: A simulation study. {\textit{BMC Medical Research Methodology}}, 14, 1-12.

\item Fischer, D. H. (1970). {\textit{Historians' fallacies: Toward a logic of historical thought}}. Harper and Row, Publishers.

\item Fischhoff, B. (1975). Hindsight is not equal to foresight: The effect of outcome knowledge on judgment under uncertainty. {\textit{Journal of Experimental Psychology: Human perception and performance}}, 1(3), 288.

\item Frisch, R., and Waugh, F. V. (1933). Partial time regressions as compared with individual trends.  {\textit{Econometrica: Journal of the Econometric Society}}, 387-401.

\item Fox, J., Weisberg, S., Adler, D., Bates, D., Baud-Bovy, G., Ellison, S., ... and Heiberger, R. (2012). Package ‘car’. {\textit{Vienna: R Foundation for Statistical Computing}}, 16(332), 333.

\item Imai, M., Murai, C., Miyazaki, M., Okada, H., and Tomonaga, M. (2021). The contingency symmetry bias (affirming the consequent fallacy) as a prerequisite for word learning: A comparative study of pre-linguistic human infants and chimpanzees. {\textit{Cognition}}, 214, 104755.

\item Kahan, B. C., Jairath, V., Doré, C. J., and Morris, T. P. (2014). The risks and rewards of covariate adjustment in randomized trials: an assessment of 12 outcomes from 8 studies. {\textit{Trials}}, 15, 1-7.

\item Kahneman, D. (1965). Control of spurious association and the reliability of the controlled variable. {\textit{Psychological Bulletin}}, 64, 326-­‐329.

\item Kempthorne, O. (1957). {\textit{An introduction to genetic statistics}}. Wiley.

\item Kennedy, P. E. (1982). Eliminating problems caused by multicollinearity: A warning. {\textit{Journal of Economic Education}}, 62-64.

\item Kerlinger, F. N., and Pedhazur, E. J. (1973). {\textit{Multiple regression in behavioral research}}. New York: Hold, Rinehart and Winston.

\item Kirk, R. E. (2013). {\textit{Experimental design: Procedures for the behavioral sciences (4th)}}. Thousand Oaks, CA: Sage.

\item Lovell, M. C. (1963). Seasonal adjustment of economic time series and multiple regression analysis. {\textit{Journal of the American Statistical Association}}, 58(304), 993-1010.

\item Meehl, P. E. (1970). Nuisance variables and the ex post facto design. In M. Radner and S. Winokur	(Eds.),	{\textit{Minnesota Studies in the philosophy of science$:$ Vol. IV. Analyses of theories and methods of physics and psychology}} (p. 373-402). Minneapolis: University of Minnesota Press.

\item Newton, R. G., and  Spurrell, D. J. (1967). A development of multiple regression for the analysis of routine data. {\textit{Journal of the Royal Statistical Society Series C: Applied Statistics}}, 16(1), 51-64.

\item Pedhazur, E. J., and Kerlinger, F. N. (1982). {\textit{Multiple regression in behavioral research}}. Holt, Rinehart, and Winston. 

\item Pocock, S. J., Assmann, S. E., Enos, L. E., and Kasten, L. E. (2002). Subgroup analysis, covariate adjustment and baseline comparisons in clinical trial reporting: current practiceand problems. {\textit{Statistics in Medicine}}, 21(19), 2917-2930.

\item Rodamar, J. (2018). There ought to be a law! Campbell versus Goodhart, {\textit{Significance}}, p. 15 (6).

\item Sherry, A. D., Msaouel, P., McCaw, Z. R., Abi Jaoude, J., Hsu, E. J., Kouzy, R., ... and Ludmir, E. B. (2023). Prevalence and implications of significance testing for baseline covariate imbalance in randomised cancer clinical trials: The Table 1 Fallacy. {\textit{European Journal of Cancer, 194}}, 113357.

\item Tesla, N. (1934). Radio power will revolutionize the world. {\textit{Modern Mechanix and Inventions}}, 2.

\item Tversky, A., and Kahneman, D. (1974). Judgment under uncertainty: Heuristics and biases: Biases in judgments reveal some heuristics of thinking under uncertainty. {\textit{Science}}, 185(4157), 1124-1131.

\item Von Elm, E., Altman, D. G., Egger, M., Pocock, S. J., Gøtzsche, P. C., Vandenbroucke, J. P., and Strobe Initiative. (2007). The Strengthening the Reporting of Observational Studies in Epidemiology (STROBE) statement: Guidelines for reporting observational studies. {\textit{Annals of Internal Medicine}}, 147(8), 573-577.
   
\item Woolf, B. (1951). Computation and interpretation of multiple regressions. {\textit{Journal of the Royal Statistical Society: Series B (Methodological), 13(1),}} 100-119.

\item Wysocki, A. C., Lawson, K. M., and Rhemtulla, M. (2022). Statistical control requires causal justification. {\textit{Advances in Methods and Practices in Psychological Science}}, 5(2), 25152459221095823.

\item Yule, G. U. (1907). On the theory of correlation for any number of variables, treated by a new system of notation. {\textit{Proceedings of the Royal Society of London. Series A, Containing Papers of a Mathematical and Physical Character}}, 79(529), 182-193.

\end{enumerate}

\newpage

\end{document}